\newcommand\apjcls{1}
\newcommand\aastexcls{2}
\newcommand\othercls{3}

\newcommand\papercls{\aastexcls}
\documentclass[tighten,times,twocolumn]{aastex62}  

\if\papercls \apjcls
\usepackage{apjfonts}
\else\if\papercls \othercls
\usepackage{epsfig}
\usepackage{margin}
\usepackage{times}
\fi\fi
\usepackage{ifthen}
\usepackage{natbib}
\usepackage{bm}
\usepackage{amssymb, amsmath}
\usepackage{appendix}
\usepackage{etoolbox}
\usepackage[T1]{fontenc}
\usepackage{paralist}
\usepackage{newtxtext,newtxmath}
\if\papercls \apjcls
\newcommand\aas{\ref@jnl{AAS Meeting Abstracts}}
\newcommand\dps{\ref@jnl{AAS/DPS Meeting Abstracts}}
\newcommand\maps{\ref@jnl{MAPS}}
\else\if\papercls \othercls
\usepackage{astjnlabbrev-jh}
\fi\fi

\def\CZdepth{
\begin{figure*}[ht!]
\begin{minipage}{\textwidth}
\includegraphics[width=1.0\textwidth]{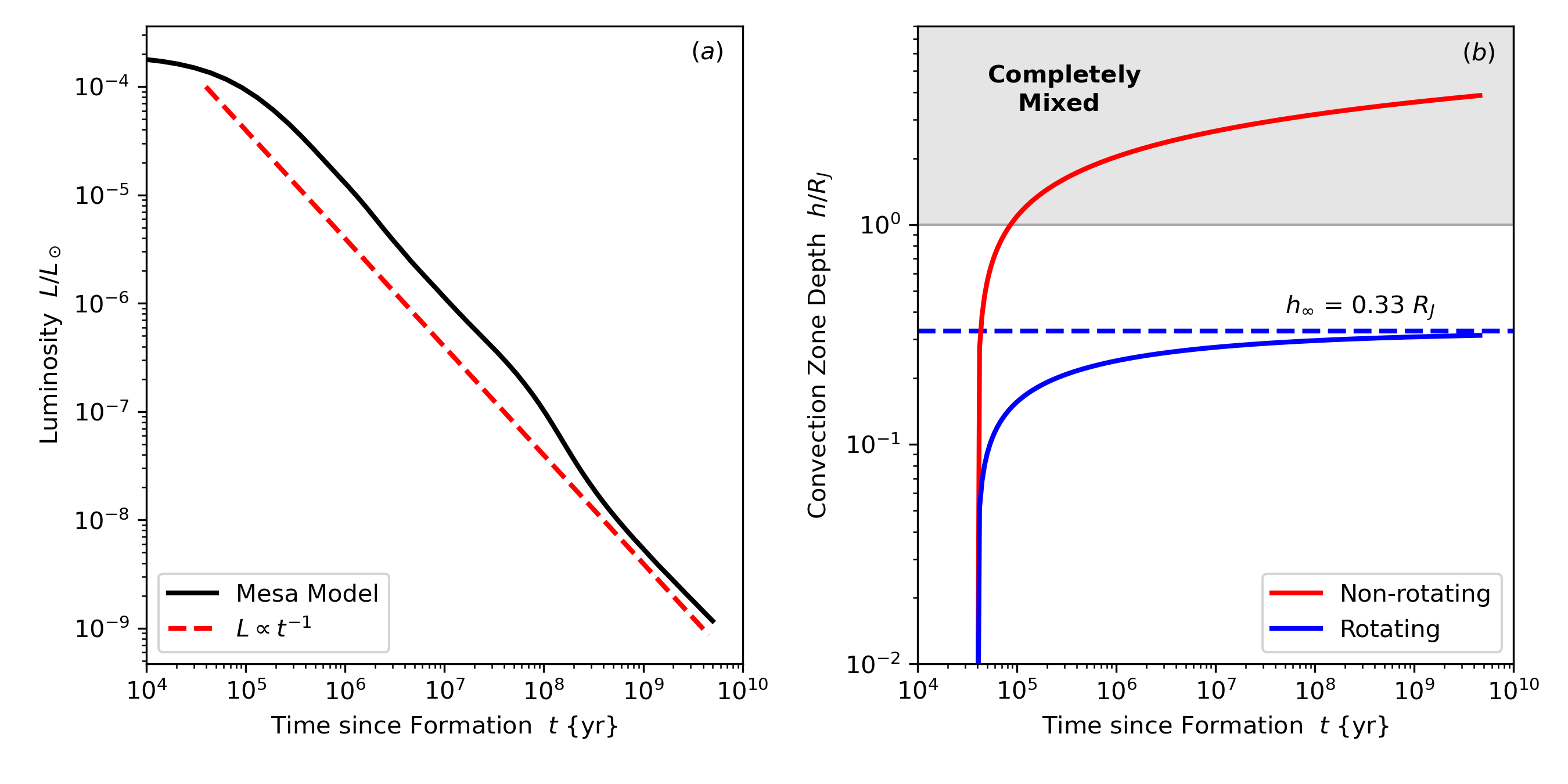}
\caption{\footnotesize Panel (a): Radiant luminosity $L$ as a function of time for a MESA model of a Jupiter-like planet \citep{Marleau2014} that undergoes cooling for 5 Gyr (black solid line) and for an analytic model where $L\propto t^{-1}$ (dashed red line).  Panel (b): Depth of the convection zone of Jupiter as a function of time for the analytic models described by Equations~\eqnref{eq:h_NR} and \eqnref{eq:h_R} and using the parameter values that appear in Table~\ref{tab:parameters}. The convection zone in a nonrotating Jupiter deepens rapidly engulfing the entire interior of the planet in less than $10^5$ years. Conversely, the convection zone in a rotating giant planet initially grows, but asymptotes to a finite depth.  For parameter values appropriate for Jupiter, the convection zone stops growing \edits{after it occupies} the outer 37\% of the planetary radius.} 
\label{fig:CZdepth}
\end{minipage}
\end{figure*}
}

\def\asympdepth{
\begin{figure*}[ht!]
\includegraphics[width=1.0\textwidth]{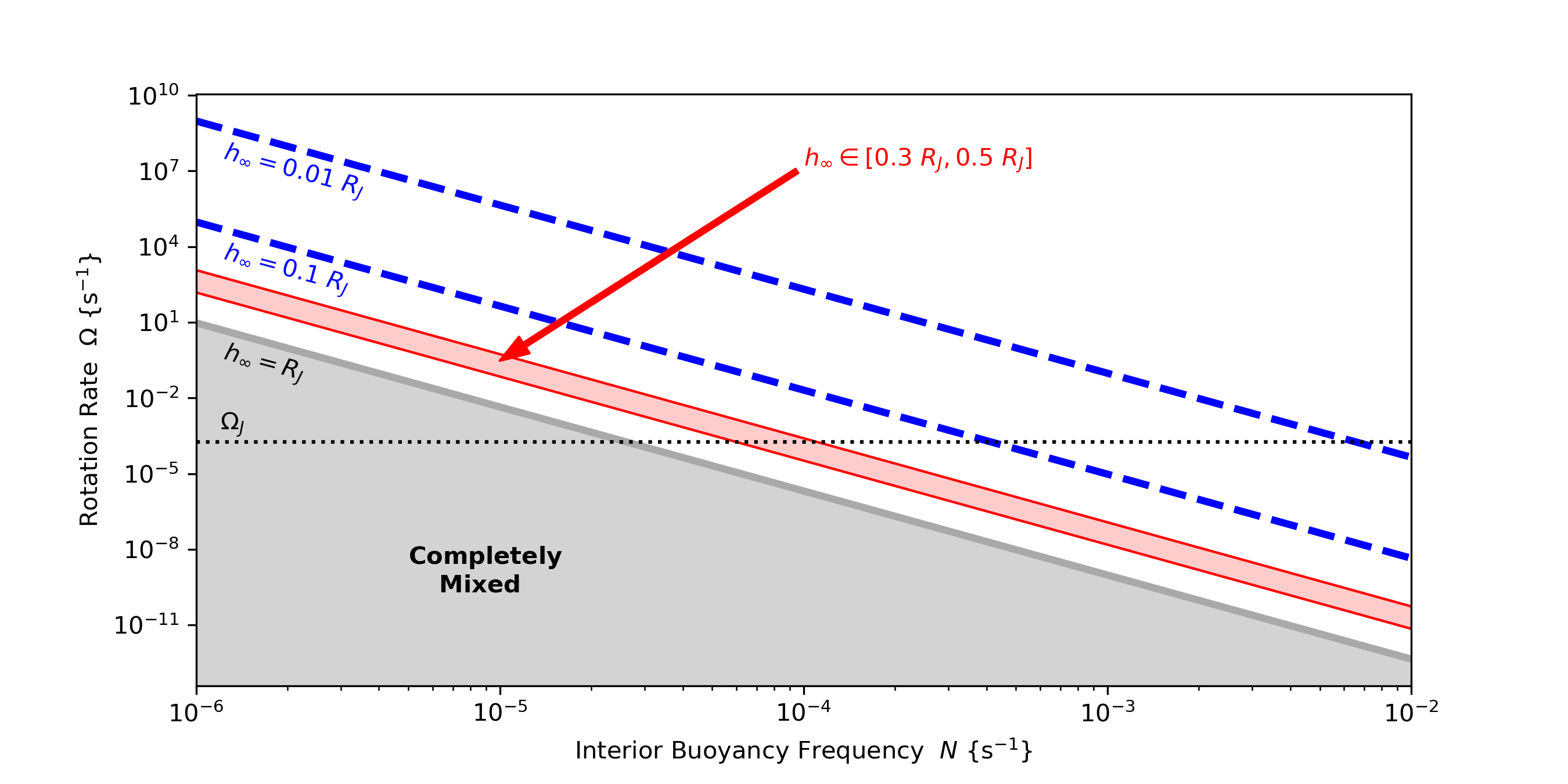}
\caption{\footnotesize  Isocountours of the asymptotic depth of the convection zone within a rotating planet with a surface cooling rate that varies like $\sim t^{-1}$. The asymptotic depth is shown as a function of the buoyancy frequency $N$ of the stably stratified interior and the planet's rotation rate $\Omega$. To ease interpretation, we have indicating the rotation rate of Jupiter with the dotted horizontal line. The two blue dashed lines provide isocontours for $h_\infty = 10^{-2}~R_J$ and for $h_\infty = 10^{-1}~R_J$.  The red shaded region indicates all values consistent with \edits{current constraints for Jupiter's outer convection zone} $h_\infty \in [0.3~R_J,0.5~R_J]$ \citep[e.g.,][]{Militzer2022}. All isocontours were calculated using Equation~\eqnref{eq:h_R} and Table~\ref{tab:parameters}.} 
\label{fig:asympdepth}
\end{figure*}
}

\if\papercls \aastexcls
\hypersetup{citecolor=blue, 
            linkcolor=blue, 
            menucolor=blue, 
            urlcolor=blue}  
\else
\usepackage[
bookmarks=true,           
bookmarksnumbered=true,   
colorlinks=true,          
citecolor=blue,           
linkcolor=blue,           
menucolor=blue,           
urlcolor=blue,            
linkbordercolor={0 0 1},  
pdfborder={0 0 1},
frenchlinks=true]{hyperref}
\fi

\if\papercls \othercls

\else

\fi

\providecommand{\adsurl}[1]{\href{#1}{ADS}}

\makeatletter
\patchcmd{\NAT@citex}
  {\@citea\NAT@hyper@{%
     \NAT@nmfmt{\NAT@nm}%
     \hyper@natlinkbreak{\NAT@aysep\NAT@spacechar}{\@citeb\@extra@b@citeb}%
     \NAT@date}}
  {\@citea\NAT@nmfmt{\NAT@nm}%
   \NAT@aysep\NAT@spacechar\NAT@hyper@{\NAT@date}}{}{}

\patchcmd{\NAT@citex}
  {\@citea\NAT@hyper@{%
     \NAT@nmfmt{\NAT@nm}%
     \hyper@natlinkbreak{\NAT@spacechar\NAT@@open\if*#1*\else#1\NAT@spacechar\fi}%
       {\@citeb\@extra@b@citeb}%
     \NAT@date}}
  {\@citea\NAT@nmfmt{\NAT@nm}%
   \NAT@spacechar\NAT@@open\if*#1*\else#1\NAT@spacechar\fi\NAT@hyper@{\NAT@date}}
  {}{}
\makeatother

\newcommand{\edits}[1]{{#1}}

\newcommand{\eps}{\varepsilon}

\newcommand{\D}{\displaystyle}
\newcommand{\eqnref}[1]{(\ref{#1})}

\begin{document}

\title{Dwindling Surface Cooling of a Rotating Jovian Planet Leads to a Convection Zone that Grows to a Finite Depth}

\author{Bradley W. Hindman}
\affil{JILA, University of Colorado, Boulder, CO~80309-0440, USA}
\affil{Department of Applied Mathematics, University of Colorado, Boulder, CO~80309-0526, USA}
\email{hindman@colorado.edu}

\author{J.R. Fuentes}
\affil{Department of Applied Mathematics, University of Colorado, Boulder, CO~80309-0526, USA}


\begin{abstract}
Recent measurements of Jupiter's gravitational field (by Juno) and seismology of Saturn's rings (by Cassini) strongly suggest that both planets have a stably-stratified core that still possesses a primordial gradient in the concentration of heavy elements. The existence of such a ``diffusely" stratified core has been a surprise as it was long expected that the Jovian planets should be fully convective and hence fully mixed. A vigorous zone of convection, driven by surface cooling, forms at the surface and deepens through entrainment of fluid from underneath.  In fact, it was believed that this convection zone should grow so rapidly that the entire planet would be consumed in less than a million years.  Here we suggest that two processes, acting in concert, present a solution to this puzzle.  All of the giant planets are rapidly rotating and have a cooling rate that declines with time. Both of these effects reduce the rate of fluid entrainment into the convection zone. Through the use of an analytic prescription of entrainment in giant planets, we demonstrate that these two effects, rotation and dwindling surface cooling, result in a convection zone which initially grows but eventually stalls.  The depth to which the convective interface asymptotes depends on the rotation rate and on the stratification of the stable interior.  Conversely, in a nonrotating planet, or in a planet that maintains a higher level of cooling than current models suggest, the convection zone deepens forever, eventually spanning the entire planet.
\end{abstract}



\section{Introduction}
\label{sec:Introduction}

Recent observations of the gravitational field of Jupiter by the Juno spacecraft and the seismology in Saturn's rings by Cassini suggest that neither planet is fully mixed. Instead of consisting of a compact segregated core of heavy elements surrounded by a deep well-mixed zone of convection \citep[as is assumed in conventional models of gas giant interiors, e.g.][]{Pollack1996}, the convection zone of each planet is shallower, occupying a half or less of the planet's radius. Below the convection zone, there likely exists a stably-stratified, diffuse core with a smoother radial gradient of heavy elements \citep{Bolton2017, Bolton2017b, Wahl2017, dc2019, Mankovich_2021, Militzer2022, Howard_2023}.

Despite the fact that composition gradients can be a natural outcome of formation models \citep[e.g.,][]{stevenson_et_al_2022}, the survival of such gradients over evolutionary timescales is not well understood. The expectation had been that at each planet's birth the high surface temperatures 
and the associated rapid cooling led to vigorous convection that quickly burrowed its way through the entirety of the planet, hence fully mixing the planet's interior on timescales as short as 1 Myr \citep[e.g., ][]{Muller2020}.

Traditionally, thermo-compositional layers (also known as ``staircases'', resembling the ones observed in the artic sea on Earth) have been proposed as a mechanism to stop the growth of the outer convection zone and prevent mixing in the deeper layers of gas giants \citep{Chabrier_Baraffe_2007,Vazan2015,Moll2017,Vazan2018}. Although a convective staircase is a plausible phenomena to occur in the interior of a gas giant, hydrodynamical simulations have shown that staircases do not persist over evolutionary timescales, as  multiple layers tend to merge over short timescales until a single well-mixed convective layer is left \citep{Mirouh2012,Wood2013,Garaud2018,Fuentes_2022,Garaud_2021}.

Recently, \cite{Fuentes2023} studied how a convection zone cooled from above at a fixed rate, mixes a primordial compositional gradient. In particular, they focused on how rotation modifies convective mixing at the boundary between the convection zone and the stable region below. Utilizing both 3D numerical simulations and recent scaling theory \citep{Barker2014,Aurnou2020} that provides estimates for the speed of turbulent convective motions in a rapidly rotating fluid, they showed that rotation significantly retards the advance of the base of the convection zone, thus reducing the mixing and entrainment of heavy elements. If confirmed with more realistic simulations, rotation would provide a simple alternative mechanism to prevent mixing in gas giants.  

Another possibility, that we will explore here, is that the luminosity of the Jovian planets has dwindled over time, starting at formation with a luminosity that is five orders of magnitude larger than at present \cite[e.g.,][]{Marleau2014}. Since convection in gas giants is driven by the fast cooling from the outer surface, this diminution of the luminous flux results in an ebbing of the convective entrainment. In this paper, we investigate the effect of convection driven by a cooling flux that decreases over time. Adapting the model of \citet{Fuentes2023}, we demonstrate that if the cooling flux decreases with sufficient rapidity, the growth of the convection zone can stall and the depth of the convecting layer asymptotes to a fixed value as time advances. Further, since rotation also diminishes the rate of entrainment, for a rotating planet the rate of decay of the cooling flux can be more sedate and still lead to a stall. In section~\ref{sec:Entrainment_model} we present an analytic model for entrainment of heavy fluid from the underlying region of stable stratification and in section~\ref{sec:Cooling_rate} we explore how a dwindling cooling rate modifies the depth of the planet's convection zone as a function of time. In section~\ref{sec:Discussion}, we summarize our findings and discuss the implications of our results for the Jovian planets.


\section{The Entrainment Model}
\label{sec:Entrainment_model}

We build an analytic entrainment model which describes the depth of a gas giant's convection zone as a function of time. We do so for both a rotating planet and a nonrotating planet by following the prescription of \citet{Fuentes2023}. However, here we allow the surface cooling rate to vary with time under the implicit assumption that the time scale for change in the cooling rate is on a long evolutionary time scale that is much longer than the convective overturning time. Hence, the convection is always in a state of statistical quasi-equilibrium, where the heat flux and other properties of the convection are allowed to equilibrate as the cooling flux evolves. This condition is easily met; for example, in Jupiter, the convective turnover time $\tau_c = h/U$ has a typical value of a year \citep{Fuentes2023} and the cooling rate changes on a time scale of a million years or longer \citep{Marleau2014}.


\subsection{Initial Atmosphere}
\edits{For simplicity, consider a plane-parallel atmosphere for which the mass density $\rho$ is a linearly increasing function of depth. A portion of the density gradient is due to vertical variation in the atmosphere's composition of heavy elements and the remainder arises from thermal stratification. We write this linear relation in the form,}

\begin{equation}
\label{eqn:initial_density}
    \rho(z) = \rho_0 \left(1 - \Gamma z \right) \;, 
\end{equation}

\noindent \edits{where $z$ is the height within the atmosphere (with $z=0$ corresponding to the top of the atmosphere) and $\rho_0$ and $\Gamma$ are positive constants that represent characteristic values of the density and the reciprocal of the density scale height}.

\subsection{Entrainment}

Surface cooling will cause a convection zone to form at the upper surface and this zone will deepen with time as convection scours the interface between the convection zone and the stably stratified fluid below. Heavy fluid will be dredged upward and mixed into the convection zone. If we assume that the convection zone is well-mixed to an adiabatic density gradient, $\rho_0 \Gamma_{\rm ad}$, and at a given time has a depth of $h(t)$, 
the change in the gravitational potential energy from the initially unmixed state, $\Delta E$, is given by


\begin{equation}
    \Delta E = \rho_0 N^2 \frac{h^3}{12} \; ,
\end{equation}

\noindent \edits{where $N$ is the buoyancy frequency, which for small density fluctuations about the fiducial density $\rho_0$ is a constant value given by $N^2 = g \left(\Gamma - \Gamma_{\rm ad}\right)$ where $g$ is the gravitational acceleration (assumed constant).}


The entrainment hypothesis states that the rate of change of potential energy is proportional to the kinetic energy flux within the convective motions \citep{Linden1975}. Dimensional analysis dictates that the kinetic energy flux is proportional to the convective flow speed, $U$, times the kinetic energy density, $\rho_0 U^2 / 2$. Hence, we find a relationship between the speed of the convective motions and the rate at which the convective interface descends,


\begin{equation}
    \label{eqn:entrainment}
   \frac{d \,\Delta E}{dt} = \rho_0 N^2 \frac{h^2}{4} \frac{dh}{dt} \approx \frac{\varepsilon}{2} \rho_0 U^3 \; ,
\end{equation}

\noindent where $\varepsilon$ is a constant of proportionality called the mixing efficiency.  Equation~\eqnref{eqn:entrainment} has been well validated by both laboratory experiments and numerical simulations \citep[e.g.,][]{Turner1968, Fernando1987, molemaker97, Fuentes2020, Fuentes2023}. \edits{The mixing efficiency $\varepsilon$ depends on the strength of turbulence of the flow, ranging from 0.1 in experiments with salty water, to approximately 1 in astrophysical flows \citep{Fuentes2020}. Since fluid motions in gas giants are highly turbulent (low viscosity), we adopt $\varepsilon = 1$.}

In a nonrotating system, the convective flow speed, $U$, can be estimated by using mixing length arguments. We start by making three assumptions: 1) the convective kinetic energy arises from buoyant acceleration over the entire depth of the convecting layer, 2) the density fluctuations $\delta\rho$ within the convection are proportional to the thermal perturbations $\delta T$, and 3) the convective heat flux equals the rate of surface cooling $F$. These three assumptions lead to the following three relations

\begin{eqnarray}
    \rho_0 U^2 &\sim& g \, \delta\rho \, h \;,
\\
    \delta \rho &\sim& \rho_0 \, \alpha \, \delta T \; ,
\\
    F &\sim& \rho_0 c_p \, \delta T \, U \; ,   
\end{eqnarray}

\noindent where the constant $c_p$ and $\alpha$ are the the specific heat capacity at constant pressure and the coefficient of thermal expansion, respectively.
When these three equations are combined, one finds that the convective velocity in the nonrotating system, $U_{\rm NR}$, scales with the cube root of the cooling rate,

\begin{equation}
    \label{eqn:Uscale_NR}
    U_{\rm NR} \sim \left(\frac{g\alpha F h}{\rho_0 c_p}\right)^{1/3} \; .
\end{equation}

When convection occurs in a rotating system, the importance of rotation is quantified by the Rossby number $\mathrm{Ro}$, defined as the ratio of the rotational period to the convective turnover time. In a rapidly rotating system, one with $\mathrm{Ro}\ll 1$, mixing-length theory leads to a very different scaling. Instead of a pure balance between inertia and buoyancy, one expects CIA balance \citep[e.g.,][]{Stevenson1979, Barker2014, Aurnou2020}, which is a three way balance between the Coriolis, inertial, and buoyancy (Archimedean) forces. This leads to a convective velocity, $U_{\rm R}$, that is significantly reduced compared to a nonrotating system,

\begin{equation}
    U_{\rm R} \sim \left(\frac{g\alpha F}{\rho_0 c_p}\right)^{2/5} \left(\frac{h}{2\Omega}\right)^{1/5}\; .
\end{equation}

\noindent In both Jupiter and Saturn ${\rm Ro} \sim 10^{-6}$, so we expect CIA balance to hold and convection to be strongly constrained by rotation.

By inserting these expressions for the convective velocity into Equation~\eqnref{eqn:entrainment}, we obtain ODEs that relate the depth of the convective layer to the cooling rate. For a nonrotating planet, we derive


\begin{equation}
\label{eqn:ODE_NR}
    \frac{\rho_0 N^2}{4\eps} \frac{c_p}{g\alpha} \, \frac{d}{dt}h_{\rm R}^2 = F \; ,
\end{equation}

\noindent whereas, for a rotating planet, we obtain


\begin{equation}
\label{eqn:ODE_R}
    \frac{5N^2}{24 \eps} \left(\frac{\rho_0 c_p}{g\alpha}\right)^{6/5} (2\Omega)^{3/5} \, \frac{d}{dt}h_{\rm R}^{12/5} = F^{6/5} \; .
\end{equation}


\section{Evolution of the Cooling Rate}
\label{sec:Cooling_rate}

When first formed, a gas giant is extremely hot and cools rapidly through radiation. As the planet ages, its surface temperature falls and the cooling rate slows. Figure~\ref{fig:CZdepth}$a$ illustrates the radiant luminosity of Jupiter as a function of time, as calculated by \citet{Marleau2014}\footnote{We use the data that is publicly available in the Github repository \href{https://github.com/andrewcumming/mesa_gasgiant}{\texttt{https://github.com/andrewcumming/mesa\_gasgiant}}}  using the MESA stellar evolutionary code \citep{Paxton2011, Paxton2013}. As noted by \citet{Marleau2014}, the luminosity dwindles at a rate roughly consistent with the reciprocal of time, i.e., $L \propto t^{-1}$. The red dashed curve in Figure~\ref{fig:CZdepth}$a$ illustrates this power law dependence. We have chosen the constant of proportionality such that $L=8.7\times 10^{-10} \, L_\odot$ at the planet's current age $t = 4.6 \times 10^9$ years, \edits{with $L_\odot = 3.846 \times 10^{33}$ ergs s$^{-1}$.}

\edits{Over the same evolutionary timescales, the rotation rate of a gas giant changes only moderately. When young and luminous, magnetic braking spins down the planet \citep[e.g.][]{Takata1996, Batygin2018} gravitational contraction due to cooling slowly spins it up. However, after about a million years the gas giant has contracted to a density where degeneracy pressure opposes further contraction \citep{Stevenson1977}, thus, the planet's radius and rotation rate stop changing. Since, the planet's rotation rate varies by factors of order unity over the entire period after planetary formation \citep[e.g., see][]{Batygin2018}, we assume a constant rotation rate. Similarly, we ignore changes in the planetary radius.}

\CZdepth


\subsection{Constant Cooling Rate}
\label{subsec:constant_rate}

For a point of comparison, first consider a cooling rate $F$ that is temporally steady. For such cooling, Equations~\eqnref{eqn:ODE_NR} and \eqnref{eqn:ODE_R} can both be integrated analytically to provide the depth of the convection zone as a function of time \citep{Fuentes2023},

\begin{eqnarray}
    h_{\rm NR}(t) &=& \left[h_0^2 + 4 \varepsilon H^2 \, Nt\right]^{1/2},
\\
    h_{\rm R}(t) &=& \left[h_0^{12/5} + \frac{24 \varepsilon}{5} H^{12/5} \!\left(\frac{N}{2\Omega}\right)^{\!\!3/5} \!\!(Nt) \right]^{5/12}\!\!,
\\
    H &\equiv& \sqrt{\frac{g \alpha F}{\rho_0 c_p N^3}} \; ,
\end{eqnarray}

\noindent where 
\edits{$H$ is an overshooting length scale that characterizes the depth to which convective plumes penetrate locally into the stable region. The integration constant $h_0$ provides the depth of the convection zone at time $t=0$.}

In the limit of long times, for the nonrotating fluid, one recovers the well-known result that the layer grows with a square-root of time dependence. However, in a rotating fluid, the growth rate is slower; the layer advances as a power law with an index of 5/12 \citep{Fuentes2023},

\begin{eqnarray}
    h_{\rm NR}(t) &\to& (4\varepsilon)^{1/2} \, H \, (Nt)^{1/2} \; ,
\\
    h_{\rm R}(t) &\to& \left(\frac{24}{5}\varepsilon\right)^{5/12} \!\! H \!\left(\frac{N}{2\Omega}\right)^{1/4} \!\!(Nt)^{5/12} \; .
\end{eqnarray}

\noindent We emphasize that in both the nonrotating and rotating planet, the convection zone continual deepens and never reaches a finite asymptotic value. 


\subsection{Dwindling Cooling Rate}
\label{subsec:dwindling_rate}

Now consider a cooling rate that dwindles like the reciprocal of time, $F = F_0 \, (t_0/t)$.  We adopt values of $F_0$ and $t_0$ to match the red-dashed line in Figure~\ref{fig:CZdepth}$a$: $F_0 = 6.0 \times 10^8$ ergs cm$^{-2}$ s$^{-1}$ and $t_0 = 4.0 \times 10^4$ such that Jupiter has a luminous flux $L = 4\pi R_J^2\, F$ that is equal to its current intrinsic luminosity ($L = 8.7\times 10^{-10}\, L_\odot$) at its current age ($t = 4.6 \times 10^9$ years) and a luminosity of $10^{-4}\, L_\odot$ at $t=t_0$. \edits{We adopt a constant Jovian radius of $R_J = 7.15 \times 10^9$ cm.} For this power-law form for the energy flux, the ODEs, Equations~\eqnref{eqn:ODE_NR} and \eqnref{eqn:ODE_R}, can be integrated analytically giving,

\begin{eqnarray}
\label{eq:h_NR}
    h_{\rm NR}(t) &=& \left[h_0^2 + 4\varepsilon H_0^2 \, (Nt_0)\ln(t/t_0)\right]^{1/2} \!,\qquad\qquad\quad 
\\  
    h_{\rm R}(t) &=& \Biggl[h_0^{12/5} + 24 \varepsilon H_0^{12/5} (Nt_0)
\nonumber\\ \label{eq:h_R}
                    & & \qquad\qquad \left.\times \left(\frac{N}{2\Omega}\right)^{\!\!3/5}\!\! \left(1- \frac{t_0^{1/5}}{t^{1/5}}\right)\right]^{5/12} \!\!,
\end{eqnarray}

\noindent where 

\begin{equation}
    H_0 \equiv \sqrt{\frac{g \alpha F_0}{\rho_0 c_p N^3}} \; .
\end{equation}

\noindent For this case with a dwindling cooling rate, the integration constant $h_0$ has a slightly different meaning, providing the depth of the convection zone when $t = t_0$ and when $F=F_0$. 


Figure~\ref{fig:CZdepth}$b$ illustrates the depth of the convection zone as a function of time for both a nonrotating planet (red curve) and for a rotating planet (blue curve). We use parameter values that are appropriate for the interior of Jupiter, see Table~\ref{tab:parameters}. For the nonrotating planet, the convection zone quickly grows, engulfing the entire Jovian interior in less than $10^5$ years. The gray region of the diagram indicates states for which the planet is fully mixed. In the rotating planet on other hand, the convection zone initially grows rapidly but then tapers off approaching a constant asymptotic value (which is marked by the horizontal dashed blue line). \edits{If we assume that the planet starts with a very shallow convection zone, $h_0 \ll R_J$}, we find that for large times, $t \gg t_0$, the asymptotic depth has the following value,

\begin{equation}
    h_{\rm R} \to h_\infty = \left(24 \varepsilon\right)^{5/12} H_0 \left(\frac{N}{2\Omega}\right)^{1/4} \! \left(N t_0\right)^{5/12} .
\end{equation}

\begin{deluxetable*}{lcccccccch}
\tablecaption{Thermodynamic and buoyancy properties of Jupiter and Saturn (order of magnitude estimates).\label{table:smodel}
}
\tablewidth{\textwidth}
\tablehead{
\colhead{} & \colhead{$\rho_0$} & 
\colhead{$\Omega$} & \colhead{$H_T = c_p/g\alpha$} & \colhead{$N$} & \colhead{$F_0$} &
\colhead{$t_0$} & \colhead{$h_0$} & \colhead{$h_\infty$} \\
\colhead{} &  \colhead{($\rm g~cm^{-3}$)} &
\colhead{($s^{-1}$)} & \colhead{($\rm cm$)} & \colhead{($\rm s^{-1}$)} & \colhead{($\rm ergs~s^{-1}~cm^{-2}$)} &
\colhead{($\rm yr$)} & \colhead{($\rm cm$)} &
}
\startdata
Jupiter & $1.3$ & $1.8\times 10^{-4}$ & $3.5\times 10^{9}$ & $10^{-4}$ & $6\times 10^8$ & $4\times 10^4$ & $7.1\times 10^6$ & $0.33 \, R_J$ \\
Saturn  & $0.7$ & $1.6\times 10^{-4}$ & $2.9\times 10^{9}$ & $4\times 10^{-4}$ & $2\times 10^8$ & $4\times 10^4$ & $6.0\times 10^6$ & $0.12 \, R_S$ 
\enddata
\tablecomments{\edits{Volumetric mean values for the density and rotation rate come from \href{https://nssdc.gsfc.nasa.gov/planetary/factsheet/}{NASA's Planetary fact sheet}. The quantity $H_T\equiv c_p/\alpha g$ is a temperature scale height that is roughly half a planetary radius, $H_T \sim 0.5 \, R_{\rm planet}$ within the deep interior of gas giants, $H_T \equiv c_p/\alpha g \sim 0.5 R_{\rm planet}$ \citep[e.g.,][]{Stevenson1977,Helled2022}. For Jupiter's buoyancy frequency, we estimate an average value from the models of \cite{Idini2022} and \cite{Lin2023} (see also discussion in Section~\ref{subsec:dwindling_rate}). For Saturn, $N$ was estimated from Figure~1b in \citet{Mankovich_2021}. Values for $F_0$ and $t_0$ for both planets are estimated from \cite{Fortney2011} and \cite{Marleau2014} (see also the discussion in Section~\ref{subsec:dwindling_rate}). The initial size of the convection zone is always $h_0\ll R_{\rm planet}$ and does not affect the estimate for the asymptotic depth.}}\label{tab:parameters}
\end{deluxetable*}


For the parameter values indicated in Table~\ref{tab:parameters}, one obtains an asymptotic depth that is only a fraction of Jupiter's radius, $h_\infty \approx 0.33 \, R_J$. Thus, due to a combination of Jupiter's rapid rotation and its dwindling luminous flux, its convection zone will never completely engulf the planet. A central core remains stably stratified and unmixed for all time.

\edits{We note that in a realistic model of a gas giant, the buoyancy frequency $N$ will be a function of depth determined by the net stratification of the dilute core. In the analytic model that we illustrate in Figure~\ref{fig:CZdepth}$b$, we adopted a constant value of $N \sim 10^{-4}~\rm s^{-1}$ within the stably stratified fluid below the convection zone. Within the convection zone itself, $N^2 \leq 0$ by definition. This value corresponds to $N \sim 0.2 \, \omega_{\rm dyn}$, where $\omega_{\rm dyn} = (GM_J/R_J^3)^{1/2} \sim 5\times 10^{-4}~\rm s^{-1}$ is the natural frequency of the planet. The fraction of $0.2$ results by averaging the buoyancy frequency in recent models of dynamical tides in Jupiter \citep{Idini2022,Lin2023} over the interior below the convective interface at $0.7R_J$. Models using recent equations of state for hydrogen-helium mixtures at Jovian conditions \citep[e.g.,][]{Militzer2022} give values that are 3--5 times larger \citep[interestingly, similar values can be estimated for Saturn, e.g., Figure~1b in][]{Mankovich_2021}}. 

\edits{Given the current uncertainty in the value of the buoyancy frequency in the outer regions of a gas giant's primordial atmosphere, in} Figure~\ref{fig:asympdepth} we present isocontours of the asymptotic depth as a function of the buoyancy frequency $N$ of the interior stable region and the rotation rate of the planet $\Omega$. For comparison the current rotation rate of Jupiter is indicated by the horizontal dotted line. The dashed blue curves indicate isocontours for values of the asymptotic depth that are a hundredth and a tenth of Jupiter's radius: $0.01 R_J$ and $0.1 R_J$. The solid gray curve indicates where the convection zone asymptotically reaches a depth equal to Jupiter's radius and the gray region marks the parameter regime where the interior of the planet eventually becomes completely mixed. The region that is shaded red corresponds to asymptotic depths that fall within the range $h_\infty \in [0.3~R_J, 0.5~R_J]$.

\edits{It is important to note that recent studies of Jupiter's gravitational moments by the Juno mission \citep[e.g.,][]{Militzer2022} have constrained the upper convection zone to span the outer 30\% of the planet's radius (being consistent with our calculations). However, \cite{Militzer2022} also found that inner regions of the planet ($r< 0.4 R_J$) can be convective and well-mixed (presumably due to dissolution of core material into the hydrogen-rich envelope). Although we do not investigate the inclusion of an inner convective zone, the gravity data does not rule out the possibility of a core that is stable to convection}.

\asympdepth


\section{Summary and Discussion}
\label{sec:Discussion}

In a gas giant, surface cooling by radiation drives a convection zone that deepens due to the entrainment of fluid from the stably-stratified fluid that lies beneath. Given that we expect a gas giant to be formed with a radial gradient in the concentration of elements heavier than helium (with heavy elements more heavily concentrated in the center), the growing depth of the convective interface should homogenize the planet's composition by dredging up fluid rich in heavy elements and mixing it throughout the convection zone. We made a simple analytic model that describes convective boundary mixing and the downward propagation of a convective interface. This model was designed to illustrate the effects of rotation and temporally varying surface cooling, hence to avoid unnecessary complications, we assume that the fluid is nearly incompressible. Using Jupiter-like values, we have shown that in a nonrotating planet, the convection zone grows rapidly and, within less than a million years, the convection zone should have ingested the entirety of the planet's interior. In a rapidly rotating planet with surface cooling that dwindles with time, the convection zone grows and saturates to a finite value, mixing only the upper $\approx 33\%$ of the planet's radius. \edits{However, we make clear that final size of the outer convection zone is sensitive to parameters that are not well-constrained in the planet's interior, particularly to the value of the buoyancy frequency ($h_{\infty} \propto N^{-5/6}$). Therefore, our models should be viewed as illustrative instead of quantitative}.

\subsection{Why Does the Convection Zone Stop Growing?}

A simple energy argument can be used to understand how the convection zone can asymptote to a finite depth. The energy required to entrain and mix the fluid ultimately comes from the surface cooling.  Cooling at the surface causes contraction and generation of gravitational potential energy in a surface boundary layer. The fluid in this boundary layer continually rains downward converting the potential energy into kinetic energy and this kinetic energy is used to perform the work needed to entrain and mix the dense fluid from the underlying layer into the convection zone.

In a nonrotating star, the flux of kinetic energy generated by falling cool fluid is directly proportional to the total amount of thermal energy that has been radiated away, i.e., $\rho_0 U^3 \propto F$---see Equation~\eqnref{eqn:Uscale_NR}.  Hence, if the cooling rate is integrable, a finite amount of energy is available for mixing and we should expect the deepening of the convection zone to stall at a finite depth. If we consider a general power law for the temporal behavior of the cooling rate, $F \propto t^{-n}$, we see that the power law index $n$ must be greater than 1 for the total radiated energy to be finite,

\begin{equation}
    \int_{t_0}^\infty F(t) dt \propto \int_{t_0}^\infty t^{-n} dt =\left\{
    \begin{array}{ll}
        \D\frac{t_0^{1-n}}{n-1} & {\rm for } \; n > 1 \; ,
        \\
        \infty & {\rm for } \; n \leq 1 \; .
    \end{array}
    \right.
\end{equation}

\noindent A power law of $n=1$ matches evolutionary models of Jupiter's luminous flux (see Figure~\ref{fig:CZdepth}$a$).  Hence, the depth of the convection zone in a nonrotating Jupiter is likely to be unbounded. Even though the value of the asymptotic depth depends on the stratification $N$, the mixing efficiency $\eps$, and other properties of the fluid ($\rho_0, \alpha, c_p$, etc.),  whether or not an asymptotic depth {\sl exists} only depends on the temporal behavior of the cooling rate. If the cooling flux has a finite integral over all time, the convection zone will stall at a finite depth.

Since rotation inhibits the generation of kinetic energy from surface cooling, the condition in a rotating planet is subtly different. The kinetic energy flux is proportional to $F^{6/5}$ instead of simply the flux $F$. Thus, the convection zone will achieve a finite asymptotic depth if the integral of $F^{6/5}$ exists and is finite,

\begin{equation}
    \int_{t_0}^\infty F(t)^{6/5} dt \propto \left\{
    \begin{array}{ll}
        \D\frac{5\,t_0^{1-6n/5}}{6n-5} & {\rm for } \; n > 5/6 \; ,
        \\
        \infty & {\rm for } \; n \leq 5/6 \; .
    \end{array}
    \right.
\end{equation}

\noindent The criterion for a finite amount of lifting work for the rotating case is that $n > 5/6$, which is satisfied by our previous choice of $n=1$. Once again the stratification and the fluid properties do not matter. Surprisingly, beyond the fact that the planet is rotating rapidly (${\rm Ro} \ll 1$), the rotation rate also does not matter. All that matters is whether $F^{6/5}$ is temporally integrable.

\subsection{Implications for Convective Staircases}

Here we have concentrated on how rotation and temporally decreasing surface cooling reduce the entrainment of fluid across a convective interface, thus leading to a slowing of the downward advance of that interface. Both of these mechanisms (rotation and dwindling cooling rate) will reduce the speed and penetration depth of convective overshoot into the stable region below the outer convection zone. Undoubtedly, these mechanism should also be important for the existence and longevity of a convective staircase that might reside underneath the outer convection zone. Rotation has been shown to hinder double-diffusive instabilities, suppressing the formation of a convective staircase \citep{Moll2017b}. However, staircases can also form by different mechanisms, such as thermohaline-shear instabilities \citep{radko_2016,garaud_2017} and convective destabilization of the diffusive boundary layer below the interface \citep[as observed in laboratory experiments and numerical simulations][]{Turner1968,Fernando1987,Fuentes_2022}. In such cases where a staircase forms despite the rotation, rotation and dwindling cooling should make staircases more stable by reducing the chance of disruption by overshooting motions from the vigorous outer convection zone.

\vspace{1cm}

\subsection{Saturn and the Ice Giants (Uranus and Neptune)}

Although we have primarily focused on Jupiter, our findings may also be relevant for Saturn, Uranus, and Neptune. For Saturn, seismology of its rings has revealed internal gravity modes trapped within a large scale composition gradient within a region that occupies the inner half of the planet's radius or larger \citep{Fuller2014,Mankovich_2021}. Hence, Saturn too is not fully mixed. \edits{Using Equations~\eqnref{eq:h_NR} and \eqnref{eq:h_R}, and planetary parameters appropriate for Saturn (see Table~\ref{tab:parameters}), we obtain an asymptotic convection-zone depth that is 12\% of Saturn's radius, $h_\infty = 0.12 \, R_S$. The shallower depth that we derive for Saturn is largely a consequence of the fact that we employ a larger buoyancy frequency for Saturn.  The density stratification (and hence the buoyancy frequency) is more constrained in Saturn than it is in Jupiter because ring seismology provides reliable and independent estimates for the internal density profile \citep{Fuller2014,Mankovich_2021}. Even when the final size of outer convection zone is smaller in our model \citep[versus $0.3$--$0.4 R_S$, see][]{Mankovich_2021}, we remind the reader that our analytic model is qualitative. }

For Uranus and Neptune, fully-mixed adiabatic models have difficulties explaining their observed luminosities \citep{Fortney2011,Nettelmann2013,Scheibe2019}, suggesting that their interiors may possess stably-stratified layers with a non-uniform distribution of heavy elements \citep{Vazan2020,Helled2011, Scheibe2021}. Moreover, incomplete radial mixing can also explain the complex magnetic fields of the two ice giants \citep{Stanley2004,Stanley2006}. If the convection zones of either Uranus or Neptune spanned the entire interior of the planet, we would expect primarily dipolar planetary magnetic fields. Thus, the fact that observed magnetic fields for the ice giants are dominated by the quadrupole contributions \citep[e.g.,][]{Connerney1987, Connerney1991}, suggests that the convection zone only occupies an outer shell. Since all of the giant planets are rapidly rotating, with thermal luminosities that decrease in time in a way qualitatively similar to Jupiter \citep{Fortney2011}, the same mechanisms that we explore here for Jupiter may result in a convection zone that occupies only a fraction of the interior of the other giant planets.

\begin{acknowledgements}
    This work was supported by NASA through grants 80NSSC18K1125, 80NSSC19K0267, and 80NSSC20K0193.
\end{acknowledgements}


\bibliography{references}{}

\begin{thebibliography}{}
\expandafter\ifx\csname natexlab\endcsname\relax\def\natexlab#1{#1}\fi

\bibitem[{{Aurnou} {et~al.}(2020){Aurnou}, {Horn}, \& {Julien}}]{Aurnou2020}
{Aurnou}, J.~M., {Horn}, S., \& {Julien}, K. 2020, PhRvR, 2, 043115

\bibitem[{{Barker} {et~al.}(2014){Barker}, {Dempsey}, \&
  {Lithwick}}]{Barker2014}
{Barker}, A.~J., {Dempsey}, A.~M., \& {Lithwick}, Y. 2014, ApJ, 791, 13

\bibitem[{{Batygin}(2018)}]{Batygin2018}
{Batygin}, K. 2018, \aj, 155, 178

\bibitem[{{Bolton} {et~al.}(2017{\natexlab{a}}){Bolton}, {Adriani},
  {Adumitroaie}, {Allison}, {Anderson}, {Atreya}, {Bloxham}, {Brown},
  {Connerney}, {DeJong}, {Folkner}, {Gautier}, {Grassi}, {Gulkis}, {Guillot},
  {Hansen}, {Hubbard}, {Iess}, {Ingersoll}, {Janssen}, {Jorgensen}, {Kaspi},
  {Levin}, {Li}, {Lunine}, {Miguel}, {Mura}, {Orton}, {Owen}, {Ravine},
  {Smith}, {Steffes}, {Stone}, {Stevenson}, {Thorne}, {Waite}, {Durante},
  {Ebert}, {Greathouse}, {Hue}, {Parisi}, {Szalay}, \& {Wilson}}]{Bolton2017}
{Bolton}, S.~J., {Adriani}, A., {Adumitroaie}, V., {et~al.} 2017{\natexlab{a}},
  Sci, 356, 821

\bibitem[{{Bolton} {et~al.}(2017{\natexlab{b}}){Bolton}, {Lunine}, {Stevenson},
  {Connerney}, {Levin}, {Owen}, {Bagenal}, {Gautier}, {Ingersoll}, {Orton},
  {Guillot}, {Hubbard}, {Bloxham}, {Coradini}, {Stephens}, {Mokashi}, {Thorne},
  \& {Thorpe}}]{Bolton2017b}
{Bolton}, S.~J., {Lunine}, J., {Stevenson}, D., {et~al.} 2017{\natexlab{b}},
  SSRv, 213, 5

\bibitem[{{Chabrier} \& {Baraffe}(2007)}]{Chabrier_Baraffe_2007}
{Chabrier}, G., \& {Baraffe}, I. 2007, ApJL, 661, L81

\bibitem[{{Connerney} {et~al.}(1987){Connerney}, {Acuna}, \&
  {Ness}}]{Connerney1987}
{Connerney}, J.~E.~P., {Acuna}, M.~H., \& {Ness}, N.~F. 1987, \jgr, 92, 15329

\bibitem[{Connerney {et~al.}(1991)Connerney, Acuña, \& Ness}]{Connerney1991}
Connerney, J. E.~P., Acuña, M.~H., \& Ness, N.~F. 1991, Journal of Geophysical
  Research: Space Physics, 96, 19023

\bibitem[{{Debras} \& {Chabrier}(2019)}]{dc2019}
{Debras}, F., \& {Chabrier}, G. 2019, ApJ, 872, 100

\bibitem[{{Fernando}(1987)}]{Fernando1987}
{Fernando}, H.~J.~S. 1987, JFM, 182, 525

\bibitem[{{Fortney} {et~al.}(2011){Fortney}, {Ikoma}, {Nettelmann}, {Guillot},
  \& {Marley}}]{Fortney2011}
{Fortney}, J.~J., {Ikoma}, M., {Nettelmann}, N., {Guillot}, T., \& {Marley},
  M.~S. 2011, ApJ, 729, 32

\bibitem[{{Fuentes} {et~al.}(2023){Fuentes}, {Anders}, {Cumming}, \&
  {Hindman}}]{Fuentes2023}
{Fuentes}, J.~R., {Anders}, E.~H., {Cumming}, A., \& {Hindman}, B.~W. 2023,
  \apjl, 950, L4

\bibitem[{{Fuentes} \& {Cumming}(2020)}]{Fuentes2020}
{Fuentes}, J.~R., \& {Cumming}, A. 2020, PhRvF, 5, 124501

\bibitem[{{Fuentes} {et~al.}(2022){Fuentes}, {Cumming}, \&
  {Anders}}]{Fuentes_2022}
{Fuentes}, J.~R., {Cumming}, A., \& {Anders}, E.~H. 2022, PhRvF, 7, 124501

\bibitem[{{Fuller}(2014)}]{Fuller2014}
{Fuller}, J. 2014, Icar, 242, 283

\bibitem[{Garaud(2017)}]{garaud_2017}
Garaud, P. 2017, JFM, 812, 1–4

\bibitem[{{Garaud}(2018)}]{Garaud2018}
{Garaud}, P. 2018, AnRFM, 50, 275

\bibitem[{{Garaud}(2021)}]{Garaud_2021}
---. 2021, arXiv e-prints, arXiv:2103.08072

\bibitem[{{Helled} {et~al.}(2011){Helled}, {Anderson}, {Podolak}, \&
  {Schubert}}]{Helled2011}
{Helled}, R., {Anderson}, J.~D., {Podolak}, M., \& {Schubert}, G. 2011, ApJ,
  726, 15

\bibitem[{{Helled} {et~al.}(2022){Helled}, {Stevenson}, {Lunine}, {Bolton},
  {Nettelmann}, {Atreya}, {Guillot}, {Militzer}, {Miguel}, \&
  {Hubbard}}]{Helled2022}
{Helled}, R., {Stevenson}, D.~J., {Lunine}, J.~I., {et~al.} 2022, Icar, 378,
  114937

\bibitem[{{Howard} {et~al.}(2023){Howard}, {Guillot}, {Bazot}, {Miguel},
  {Stevenson}, {Galanti}, {Kaspi}, {Hubbard}, {Militzer}, {Helled},
  {Nettelmann}, {Idini}, \& {Bolton}}]{Howard_2023}
{Howard}, S., {Guillot}, T., {Bazot}, M., {et~al.} 2023, A\&A, 672, A33

\bibitem[{{Idini} \& {Stevenson}(2022)}]{Idini2022}
{Idini}, B., \& {Stevenson}, D.~J. 2022, PSJ, 3, 89

\bibitem[{{Lin}(2023)}]{Lin2023}
{Lin}, Y. 2023, A\&A, 671, A37

\bibitem[{{Linden}(1975)}]{Linden1975}
{Linden}, P.~F. 1975, Journal of Fluid Mechanics, 71, 385

\bibitem[{{Mankovich} \& {Fuller}(2021)}]{Mankovich_2021}
{Mankovich}, C.~R., \& {Fuller}, J. 2021, NatAs, 5, 1103

\bibitem[{{Marleau} \& {Cumming}(2014)}]{Marleau2014}
{Marleau}, G.~D., \& {Cumming}, A. 2014, MNRAS, 437, 1378

\bibitem[{{Militzer} {et~al.}(2022){Militzer}, {Hubbard}, {Wahl}, {Lunine},
  {Galanti}, {Kaspi}, {Miguel}, {Guillot}, {Moore}, {Parisi}, {Connerney},
  {Helled}, {Cao}, {Mankovich}, {Stevenson}, {Park}, {Wong}, {Atreya},
  {Anderson}, \& {Bolton}}]{Militzer2022}
{Militzer}, B., {Hubbard}, W.~B., {Wahl}, S., {et~al.} 2022, PSJ, 3, 185

\bibitem[{{Mirouh} {et~al.}(2012){Mirouh}, {Garaud}, {Stellmach}, {Traxler}, \&
  {Wood}}]{Mirouh2012}
{Mirouh}, G.~M., {Garaud}, P., {Stellmach}, S., {Traxler}, A.~L., \& {Wood},
  T.~S. 2012, ApJ, 750, 61

\bibitem[{Molemaker \& Dijkstra(1997)}]{molemaker97}
Molemaker, M.~J., \& Dijkstra, H.~A. 1997, JFM, 331, 199–229

\bibitem[{{Moll} \& {Garaud}(2017)}]{Moll2017b}
{Moll}, R., \& {Garaud}, P. 2017, ApJ, 834, 44

\bibitem[{{Moll} {et~al.}(2017){Moll}, {Garaud}, {Mankovich}, \&
  {Fortney}}]{Moll2017}
{Moll}, R., {Garaud}, P., {Mankovich}, C., \& {Fortney}, J.~J. 2017, ApJ, 849,
  24

\bibitem[{{M{\"u}ller} {et~al.}(2020){M{\"u}ller}, {Helled}, \&
  {Cumming}}]{Muller2020}
{M{\"u}ller}, S., {Helled}, R., \& {Cumming}, A. 2020, A\&A, 638, A121

\bibitem[{{Nettelmann} {et~al.}(2013){Nettelmann}, {Helled}, {Fortney}, \&
  {Redmer}}]{Nettelmann2013}
{Nettelmann}, N., {Helled}, R., {Fortney}, J.~J., \& {Redmer}, R. 2013, P\&SS,
  77, 143

\bibitem[{{Paxton} {et~al.}(2011){Paxton}, {Bildsten}, {Dotter}, {Herwig},
  {Lesaffre}, \& {Timmes}}]{Paxton2011}
{Paxton}, B., {Bildsten}, L., {Dotter}, A., {et~al.} 2011, \apjs, 192, 3

\bibitem[{{Paxton} {et~al.}(2013){Paxton}, {Cantiello}, {Arras}, {Bildsten},
  {Brown}, {Dotter}, {Mankovich}, {Montgomery}, {Stello}, {Timmes}, \&
  {Townsend}}]{Paxton2013}
{Paxton}, B., {Cantiello}, M., {Arras}, P., {et~al.} 2013, \apjs, 208, 4

\bibitem[{{Pollack} {et~al.}(1996){Pollack}, {Hubickyj}, {Bodenheimer},
  {Lissauer}, {Podolak}, \& {Greenzweig}}]{Pollack1996}
{Pollack}, J.~B., {Hubickyj}, O., {Bodenheimer}, P., {et~al.} 1996, Icar, 124,
  62

\bibitem[{Radko(2016)}]{radko_2016}
Radko, T. 2016, JFM, 805, 147–170

\bibitem[{{Scheibe} {et~al.}(2019){Scheibe}, {Nettelmann}, \&
  {Redmer}}]{Scheibe2019}
{Scheibe}, L., {Nettelmann}, N., \& {Redmer}, R. 2019, A\&A, 632, A70

\bibitem[{{Scheibe} {et~al.}(2021){Scheibe}, {Nettelmann}, \&
  {Redmer}}]{Scheibe2021}
---. 2021, A\&A, 650, A200

\bibitem[{{Stanley} \& {Bloxham}(2004)}]{Stanley2004}
{Stanley}, S., \& {Bloxham}, J. 2004, Natur, 428, 151

\bibitem[{{Stanley} \& {Bloxham}(2006)}]{Stanley2006}
---. 2006, Icar, 184, 556

\bibitem[{{Stevenson}(1979)}]{Stevenson1979}
{Stevenson}, D.~J. 1979, GApFD, 12, 139

\bibitem[{{Stevenson} {et~al.}(2022){Stevenson}, {Bodenheimer}, {Lissauer}, \&
  {D'Angelo}}]{stevenson_et_al_2022}
{Stevenson}, D.~J., {Bodenheimer}, P., {Lissauer}, J.~J., \& {D'Angelo}, G.
  2022, PSJ, 3, 74

\bibitem[{{Stevenson} \& {Salpeter}(1977)}]{Stevenson1977}
{Stevenson}, D.~J., \& {Salpeter}, E.~E. 1977, \apjs, 35, 239

\bibitem[{{Takata} \& {Stevenson}(1996)}]{Takata1996}
{Takata}, T., \& {Stevenson}, D.~J. 1996, \icarus, 123, 404

\bibitem[{{Turner}(1968)}]{Turner1968}
{Turner}, J.~S. 1968, JFM, 33, 183

\bibitem[{{Vazan} \& {Helled}(2020)}]{Vazan2020}
{Vazan}, A., \& {Helled}, R. 2020, A\&A, 633, A50

\bibitem[{{Vazan} {et~al.}(2018){Vazan}, {Helled}, \& {Guillot}}]{Vazan2018}
{Vazan}, A., {Helled}, R., \& {Guillot}, T. 2018, A\&A, 610, L14

\bibitem[{{Vazan} {et~al.}(2015){Vazan}, {Helled}, {Kovetz}, \&
  {Podolak}}]{Vazan2015}
{Vazan}, A., {Helled}, R., {Kovetz}, A., \& {Podolak}, M. 2015, ApJ, 803, 32

\bibitem[{{Wahl} {et~al.}(2017){Wahl}, {Hubbard}, {Militzer}, {Guillot},
  {Miguel}, {Movshovitz}, {Kaspi}, {Helled}, {Reese}, {Galanti}, {Levin},
  {Connerney}, \& {Bolton}}]{Wahl2017}
{Wahl}, S.~M., {Hubbard}, W.~B., {Militzer}, B., {et~al.} 2017, GeoRL, 44, 4649

\bibitem[{{Wood} {et~al.}(2013){Wood}, {Garaud}, \& {Stellmach}}]{Wood2013}
{Wood}, T.~S., {Garaud}, P., \& {Stellmach}, S. 2013, ApJ, 768, 157

\end{thebibliography}
\bibliographystyle{aasjournal}

\end{document}